\documentclass[%
 reprint,
superscriptaddress,
amsmath,amssymb,
prl,
]{revtex4-1}

\usepackage{graphicx}
\usepackage{dcolumn}
\usepackage{bm}

\begin{document}

\preprint{APS/123-QED}

\title{Evidence for two-electron processes in the mutual neutralization \\ of O$^-$ with O$^+$ and N$^+$ at Subthermal Collision Energies}%

\author{N. de Ruette}
\affiliation{%
 Department of Physics, Stockholm University, Stockholm, SE-106 91, Sweden
}%
\affiliation{%
 Columbia Astrophysics Laboratory, Columbia University, New York, NY 10027, U.S.A.
}%

\author{A.~Dochain}
\affiliation{
 Institute of Condensed Matter and Nanosciences, Universit\'e catholique de Louvain, B-1348 Louvain-la-Neuve, Belgium
}%

\author{T. Launoy}
\affiliation{
 Laboratoire de Chimie Quantique et Photophysique, Universit\'e Libre de Bruxelles, B-1050 Brussels, Belgium
}

\author{R. F. Nascimento}
\affiliation{%
 Centro Federal de Educa\c c\~ao Tecnol\'ogica Celso Suckow da Fonseca, Petr\'opolis, 25620-003, RJ, Brazil​
}%
\affiliation{%
 Department of Physics, Stockholm University, Stockholm, SE-106 91, Sweden
}%

\author{M. Kaminska}
\affiliation{%
 Department of Physics, Stockholm University, Stockholm, SE-106 91, Sweden
}%
\affiliation{%
 Institute of Physics, Jan Kochanowski University, 25-369 Kielce, Poland
}%

\author{M. H. Stockett}
\affiliation{%
 Department of Physics, Stockholm University, Stockholm, SE-106 91, Sweden
}%

\author{N. Vaeck}
\affiliation{
	Laboratoire de Chimie Quantique et Photophysique, Universit\'e Libre de Bruxelles, B-1050 Brussels, Belgium
}

\author{H. T. Schmidt} 

\author{H. Cederquist}
\affiliation{%
 Department of Physics, Stockholm University, Stockholm, SE-106 91, Sweden
}%


\author{X. Urbain}
\affiliation{
 Institute of Condensed Matter and Nanosciences, Universit\'e catholique de Louvain, B-1348 Louvain-la-Neuve, Belgium
}%

\date{\today}

\begin{abstract}
We have measured total absolute cross sections for the Mutual Neutralization (MN) of O$^-$ with O$^+$/N$^+$. A fine resolution (of about 50 meV) in the kinetic energy spectra of the product neutral atoms allows unique identification of the atomic states participating in the mutual neutralization process. Cross sections and branching ratios have also been
calculated down to 1 meV center-of-mass collision energy for these two systems with a multi-channel Landau-Zener model and an asymptotic method for the ionic-covalent coupling matrix elements. The importance of two-electron processes in one-electron transfer is demonstrated by the dominant contribution of a core-excited configuration of the nitrogen atom in N$^+$ + O$^-$ collisions. This effect is partially accounted for by introducing configuration mixing in the evaluation of coupling matrix elements. 

\end{abstract}

\pacs{Valid PACS appear here} 
\maketitle

Anions play crucial roles in a range of astrophysical environments and planetary atmospheres. Atomic and molecular negative ions, such as O$^-$ and O$_2^-$, have been detected in the atmospheres of Earth \cite{Smit95,Chutjian}, Mars \cite{Larsson}, and Titan \cite{Vuitton}. Already in 1939, Wildt \cite{Wildt} suggested that the simplest atomic anion H$^-$ is responsible for the opacity of the solar photosphere at wavelengths below 1645~nm, and shortly thereafter this was verified through direct spectral observations \cite{Rau,Chandrasekhar,Massey40}. It was also suggested that H$^-$ \cite{Massey50,Chutjian} and other anions such as C$^-$, Cl$^-$, S$^-$, O$^-$ and O$_2^-$ \cite{Branscomb,Vardya} influence stellar absorption spectra in general, although the concentrations of O$^-$ and O$_2^-$ turned out to be too small to be detected in these environments. Reactions with anions have been included in interstellar chemistry models \cite{DalgMcCray,Herbst81} for decades, but it was not until 2006 that the first negative molecular ion, C$_6$H$^-$, was observed in molecular clouds~\cite{McCarthy}. Since then C$_4$H$^-$~\cite{Cernicharo2007}, C$_8$H$^-$~\cite{Brunken}, C$_3$N$^-$~\cite{Thaddeus}, C$_5$N$^-$~\cite{Cernicharo2008}, and most recently CN$^-$~\cite{Agundez}, have also been detected. Negative ions such as O$^-$ and/or OH$^-$ and C$^-$ and/or CH$^-$, were detected with limited mass resolution in the coma of comet 1P/Halley \cite{Chaizy}. Further it was proposed that reactions involving O$^-$, S$^-$ and C$^-$ could be responsible for the presence of CS molecules in the supernova 1987A ejecta~\cite{liu1998}.

Anions often have loosely bound outer electrons and large reactivities, and may thus influence the charge balance in plasma even at low concentrations. Therefore atomic anion formation processes such as dissociative ($\rm AB + e^- \rightarrow A^- + B$) and radiative ($\rm A + e^- \rightarrow A^- + h\nu$) attachment are of large interest for, e.g. astrophysical modeling \cite{Larsson,Harada}. Likewise anion {\em destruction} processes such as photodetachment ($\rm A^- + h\nu \rightarrow A + e^-$), associative detachment ($\rm A^- + B \rightarrow AB + e^-$), associative ionisation (AI) ($ \rm A^- + \rm B^+ \rightarrow \rm AB^+ + e^-$) and mutual neutralization~(MN) 
\begin{eqnarray}
\rm A^- + \rm B^+ \rightarrow \rm A + B  \label{Eq:MN}
\end{eqnarray} 	
are important in such contexts. The corresponding absolute cross sections and rate coefficients are thus key to determine charge balances, electron concentrations, and abundances of different atomic and molecular species (in the forms of neutrals and ions) in cold molecular clouds and in other astrophysical environments \cite{Larsson,Flower,Cordiner12,Millar,Harada,Wake08,Walsh,Wake12a,UMIST13,Cordiner}.  

Earlier experiments on mutual neutralization have mostly been performed at collision energies above some tenths of an eV. This energy range is not the most relevant for many astrophysical applications and, further, different final quantum states were not resolved in these earlier studies \cite{Hayton,Terao}. 
With the present experimental technique, two co-propagating atomic ion beams are merged such that the angles between individual anion and cation trajectories typically are smaller than 1.5~mrad in the interaction region \cite{Nkambule2016}. We are thus able to present the first subthermal studies of quantum-state resolved MN processes. We report results for the O$^- + \rm O^+$ and O$^- + \rm N^+$ MN reactions for center-of-mass collision energies between 5 meV and 2 eV. The mutual neutralization experiments were performed at the Universit\'e Catholique de Louvain. Studies of associative ionization AI processes were previously performed with the same apparatus \cite{Nzeyi} and the corresponding results may be used to put the MN reaction cross sections on an absolute scale. 

\begin{figure}[tbp]
	\centering
	\includegraphics[width=\columnwidth]{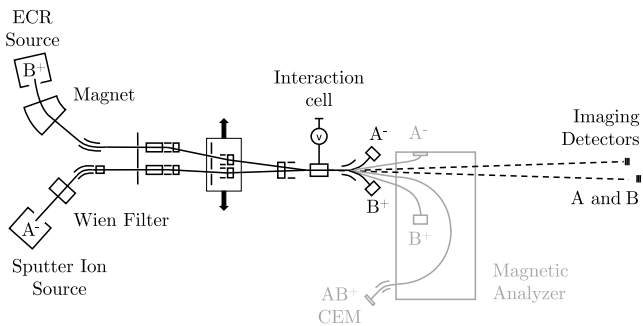} 
	\caption{Experimental apparatus (not to scale). The gray elements are used for Associative Ionization measurements.}
	\label{Fig:apparatus}
\end{figure}

The negative ion beam O$^-$ is produced by a Cs-ion sputter source from a Fe$_2$O$_3$ cathode. A Wien filter is used to select the ion mass. An electron cyclotron resonance (ECR) ion source followed by a bending magnet is used to form the O$^+$ and the N$^+$ beams. After the beams are shaped with ion optics and collimators \cite{Nzeyi}, they are merged in a 6.8$\pm$0.2~cm long interaction cell. The voltage on this cell can be fine-tuned in order to adjust the center-of-mass collision energy down to the meV range, where the lower limit is set by the angular spread within each of the ion beams and any (small) misalignment between them. The beams are demerged before leaving the ultrahigh vacuum section of the apparatus. A set of deflector plates after the interaction region is used to separate the ionic parent beams from the neutral products and to send the A$^-$ and B$^+$ ion beams into separate Faraday cups as shown in Fig.\ref{Fig:apparatus} (in black for MN and in gray for AI measurements). For the AI measurements, the product molecular ions AB$^+$ are deflected by a 180$^{\circ}$ magnet followed by a 30$^{\circ}$ deflector -- to filter out scattered ions -- and are sent into a single counting mode channel electron multiplier (CEM) as shown in gray in Fig. \ref{Fig:apparatus}. 

To detect the neutral products of mutual neutralization, we use a three-dimensional imaging detection system. It consists of two position sensitive detectors, each composed of a stack of three microchannel plates (MCPs) and a resistive anode. They are separated in the beam-propagation direction by 10 cm to reduce the dead zone between them. The kinetic energy release (KER) is determined from coincidence measurements of the positions and the differences in time of arrival of two neutrals (A and B) hitting separate detectors in a single MN event. The distortion on the positions caused by non-linear spatial response of the detectors is corrected for and the background due to false coincidences subtracted. The spectra are then corrected for the KER-dependent angular acceptance \cite{fabre}.
	
In contrast to earlier studies of MN \cite{Szucs}, the use of two separate detectors allows for simultaneous detection of the two products and the long drift distance of 3.25~m from the interaction cell to the imaging detectors allow us to minimize the misalignment of the two beams. This is done by optimizing the MN coincidence rate relying on the (expected) $1/E_{CM}$ energy dependence of the cross section and the fact that the angular dispersion of the beams is the main limiting factor for the resolution in the definition of the center-of-mass collision energy, $E_{CM}$ \cite{Brou83a}. Considering a collision between an anion of mass $m_A$ and kinetic energy $E_A$ and a cation of mass $m_B$ and kinetic energy $E_B$ ($E_A$ and $E_B$ are the ion-beam energies in the laboratory system), we get
\begin{equation}
E_{CM}= \mu\left(\frac{E_A}{m_A}+\frac{E_B}{m_B}-2\sqrt{\frac{E_AE_B}{m_Am_B}}\cos\phi\right) \label{ECM}
\end{equation}
where $\mu=m_Am_B/(m_A+m_B)$ is the reduced mass and $\phi$ is the angle between the ion trajectories. For 7 keV oxygen beams, $E_A/m_A=E_B/m_B$ and $\phi=1$ mrad, $E_{CM} = 3.5$ meV. A typical spread $\Delta E_A \approx 5$ eV in the anion beam energy then gives a negligible spread of $\Delta E_{CM} \approx 10^{-6}$ eV. The sensitivity to a similar energy spread in the cation beam is equally low. However, a spread in angles $\Delta \phi$ between anion and cation trajectories of 1 mrad, gives $\Delta E_{CM} \approx$ 7~meV.
   
The ratio between the finite length of the interaction region and the distance to the detectors determines the KER resolution. A longer flight distance gives a better precision on the velocity measurements of the neutrals and thus increases the resolution, but also limits the angular acceptance. Here, we reach a resolution of 50 meV FWHM at 1 eV of KER and we are thus able to identify the quantum states (L-S terms) of the neutral reaction products.

\begin{figure}[h!]
	\begin{center}
		\includegraphics[width=\columnwidth]{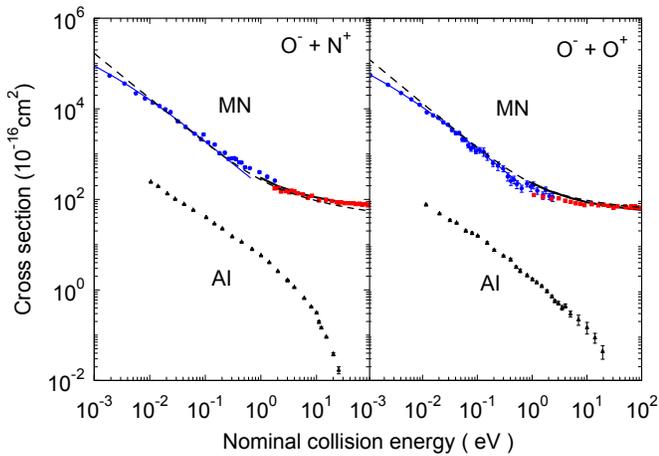} 
		\caption{Total cross sections for $\mathrm{O}^- + \mathrm{X}^+ \rightarrow \mathrm{O} + \mathrm{X}^*$ (MN, blue circles from present data; red squares from Ref. \cite{Hayton}) and for $\mathrm{O}^- + \mathrm{X}^+ \rightarrow \mathrm{XO}^+ + \mathrm{e}^-$ (AI, black triangles from Ref. \cite{Nzeyi}) as functions of the nominal collision energy ($\phi = 0$ in Eq. \ref{ECM}). Left panel: X = N, right panel: X = O. The solid black lines are calculations from Zhou \& Dickinson \cite{Zhou1997} down to 1~eV. The dashed black lines show the present calculations. The blue lines are the convolution of E$_{CM}^{-1}$ cross sections with the angular and energetic distribution of the beams (see text).}
		\label{Fig:cs}
	\end{center}
\end{figure}

Total and angular differential MN cross sections can be retrieved from the measured angular distributions. In Figure \ref{Fig:cs}, we show the present total and absolute MN cross section for $\mathrm{O}^- + \mathrm{N}^+ \rightarrow \mathrm{O} + \mathrm{N}^*$ (blue circles, top panel) and for $\mathrm{O}^- + \mathrm{O}^+ \rightarrow \mathrm{O} + \mathrm{O}^*$ (blue circles, bottom panel). The MN absolute cross section scale is established by means of previously measured absolute AI cross sections \cite{Nzeyi}, using the presently measured ratio between AI and MN rates, assuming 50 \% MCP detection efficiency and correcting for geometrical limitations. This gives good agreement with the results of Hayton and Peart \cite{Hayton} (red squares) as well as with calculations (solid black line) by Zhou and Dickinson \cite{Zhou1997} down to 1~eV.

\begin{figure}[ht!]
	\begin{center}
		\includegraphics[width=\columnwidth]{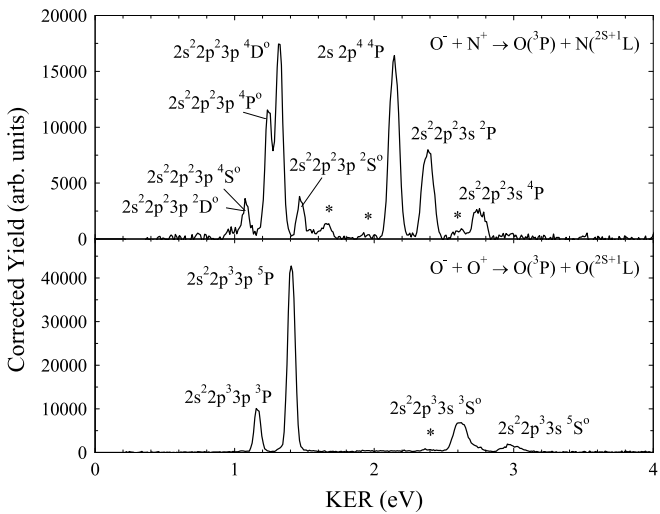} 
		\caption{Kinetic Energy Release spectra at 5 meV center-of-mass collision energy for the MN of $\rm O^- + \rm N^+$ (top panel) and $\rm O^- + \rm O^+$ (bottom panel). The stars indicate contributions from excited cations.}
		\label{Fig:spectra}
	\end{center}
\end{figure}

The KER distributions for O$^-$ colliding with N$^+$ (top panel) and with O$^+$ (bottom panel) are shown in Figure~\ref{Fig:spectra}. Each peak in these spectra corresponds to a separate L-S term in the excited neutral N (upper panel of Fig. \ref{Fig:spectra}) and in the excited O atom (lower panel) while the (other) O atom is in its ground state after it has lost its extra electron. The peaks marked with stars are due to cations in metastable excited states before the interaction. We observe a shift in the peaks' KER of $\approx$~5~meV in relation to their expected positions as given by KER~=~$IE_B~-~EA_A~-~E^{exc}_B$ with $IE_B$ being the ionization energies of the cations, $EA_A$ the electron affinity of O$^-$ and $E^{exc}_B$ the (2J+1)-weighted mean excitation energy of the statistically populated neutrals formed in electron capture by the cations. Since these spectra were measured at nominal collision energies of 0 eV according to Eq. \ref{ECM} with an assumed angle of $\phi = 0$, the shift in energy corresponds to the collision energy $E_{CM}$ and is due to the angular spread of the beams.

\begin{figure}[ht!]
	\begin{center}
		\includegraphics[width=\columnwidth]{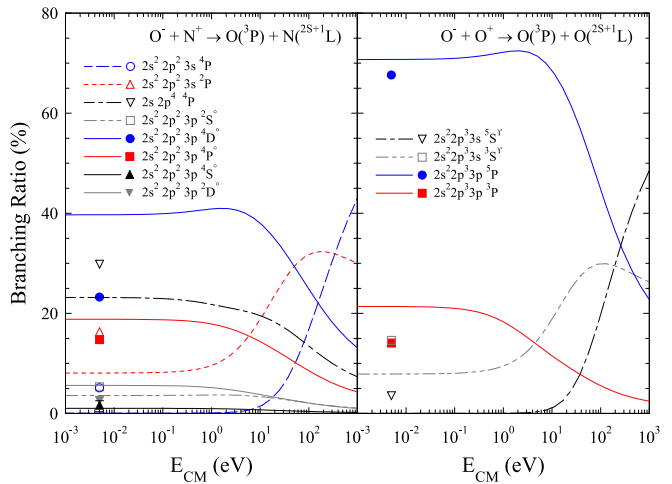} 
		\caption{Branching ratios for $\rm O^- + \rm N^+$ (left) and $\rm O^- + \rm O^+$ (right) as function of $E_{CM}$. The lines are the present calculations: left panel with MCHF coupling elements and right panel with initial values from \cite{Zhou1997}. Symbols: present experiment.}
		\label{BR-NO-OO}
	\end{center}
\end{figure}

The area of individual peaks in the KER spectra yields the branching ratios and the corresponding absolute state-selective cross section (by relating the peak area to the total intensity in the KER spectrum). The branching ratios are given in Table~\ref{BR-NO} for the O$^-$ + N$^+$ system and shown in Fig. \ref{BR-NO-OO} for both systems.

The method of Zhou and Dickinson \cite{Zhou1997}, based on a multi-channel Landau-Zener (LZ) model and the Firsov-Landau-Herring \cite{Chibisov1988} method, was used to calculate total and partial cross sections down to 1 meV collision energy for both collision systems. The asymptotic method of Firsov-Landau-Herring allows the evaluation of the one-electron exchange interaction $\Delta(R_X)$, where $R_X$ is the crossing distance, and thus the corresponding coupling matrix element $H_{if} \sim \Delta(R_X)/2$ between states $i$ and $f$. 

\begin{table}

	\begin{ruledtabular}
	\begin{tabular}{ll@{\hspace{5ex}}crrr}
	 L-S term &  & E$_{exc}$(eV) & BR(\%) exp. & ZD & MCHF \\
	\hline\\	[-1.5ex]
	2s$^2$ 2p$^2$ 3s & $^4$P & 10.3323 & 5.11 $\pm$ 0.11 & 0.05 & 0.04 \\
	2s$^2$ 2p$^2$ 3s & $^2$P & 10.6865 & 16.15 $\pm$ 0.18 & 13.35 & 8.08 \\
	2s 2p$^4$ & $^4$P & 10.9270 & 29.99 $\pm$ 0.22 & 1.09 & 23.17 \\
	2s$^2$ 2p$^2$ 3p & $^2$S$^o$ & 11.6026 & 5.34 $\pm$ 0.12 & 4.68 & 3.58 \\
	2s$^2$ 2p$^2$ 3p & $^4$D$^o$ & 11.7584 & 23.25 $\pm$ 0.19 & 50.11 & 39.69 \\
	2s$^2$ 2p$^2$ 3p & $^4$P$^o$ & 11.8417& 14.74 $\pm$ 0.16 & 22.77 & 18.82 \\
	2s$^2$ 2p$^2$ 3p & $^4$S$^o$ & 11.9956 & 1.44 $\pm$ 1.10 & 1.21 & 1.02 \\
	2s$^2$ 2p$^2$ 3p & $^2$D$^o$ & 12.0058 & 2.79 $\pm$ 0.35 & 6.74 & 5.60 \\
	\end{tabular}
	\end{ruledtabular}
		\caption{Branching ratios (in \%) for O$^-$ + N$^+$ $\rightarrow$ O($^3$P) + N($^{2s+1}$L) at 5 meV CM collision energy for each L-S term of N: statistically averaged excitation energy E$_{exc}$ in eV, experimental results (Exp.), calculations with initial values from \cite{Zhou1997}(ZD) and modified (MCHF) coupling elements.} \label{BR-NO} 
\end{table}

The coupling elements of Zhou and Dickinson \cite{Zhou1997} were first used for the calculation of state-selective cross sections. The corresponding branching ratios were then extracted from the calculated partial cross sections and are shown in Table~\ref{BR-NO} for collisions with N$^+$ (at 5 meV) and as functions of the center-of-mass energy in the right panel of Fig. \ref{BR-NO-OO} for collisions with O$^+$. For the latter system, the agreement between experimental and calculated branching ratios at 5 meV for the four states appearing in the KER spectrum is fairly good. However, for the O$^-$ + N$^+$ system, if we compare the calculations (column ZD in Table~\ref{BR-NO}) to the experiment, the 2s 2p$^4$ $^4$P state is clearly underestimated. 

In order to investigate the origin of the strong coupling to the 2s 2p$^4$ $^4$P state in N, we performed multiconfiguration Hartree-Fock (MCHF) calculations of the 2s~2p$^4$~$^4$P energy using an expansion with the three $^4$P configurations 2s 2p$^4$ $^4$P, 2s$^2$ 2p$^2$ 3s $^4$P and 2s$^2$ 2p$^2$ 3d $^4$P. This yielded the corresponding mixing coefficients and these were then used to obtain a modified coupling element for the N(2s 2p$^4$ $^4$P) state. Branching ratios using coupling strengths based on MCHF calculations are shown in the last column of Table~\ref{BR-NO} for 5~meV and in the left panel of Fig. \ref{BR-NO-OO}, as functions of center-of-mass energy. They are in better agreement with the experiment. However, all the 2s$^2$ 2p$^2$ 3p channels, i.e. the channels with low KER, are overestimated by theory while the 2s~2p$^4$~$^4$P and 2s$^2$ 2p$^2$ 3s channels (high KER) are still underestimated.

Our measurements and calculations clearly show that states, such as 2s~2p$^4$~$^4$P, populated by a two-electron process can not, as they often are, be neglected in calculations. Moreover, we can see that multi-channel LZ models underestimate high KER channels. In these approaches, only two-by-two couplings at large and intermediate distances are considered, neglecting transitions that may be active at smaller internuclear distances. This was shown by Mitrushchenkov {\em et al.} \cite{Mitrushchenkov} for the Ca$^+$~+~H$^-$ MN reaction. They compared a branching probability current method with the multichannel approach and showed that the latter underestimates the contribution of weak transitions to the cross section. This effect is important for the Ca(3d4p $^3$F) state, which is similar to N(2s 2p$^4$ $^4$P) since its population implies a two-electron process. However, the branching probability current method can not be applied to O$^-$ + N$^+$ as it is too complex for current computational capabilities. 

Crossings at short distances are inherently difficult to treat as they are not localized, may overlap with one another and are affected by the flux branching at the previous crossings along the way to short distances. A way to improve the calculations could be to combine {\em ab initio} calculations at the smaller distances with LZ calculations at larger distances. The latter gives enough accuracy for the highly excited electronic states and avoided crossings at large distances while the former is too complex in that range.

The total cross sections calculated with the coupling elements from Zhou \& Dickinson \cite{Zhou1997} for O$^-$ + O$^+$ and from coupling elements modified by means of the present MCHF calculations for O$^-$ + N$^+$, are shown by the dashed lines in Fig. \ref{Fig:cs}. They follow the expected $E_{CM}^{-1}$ trend. In both cases, the deviation from the experimental results at low collision energies is a result of the deviation of the actual collision energy from the detuning energy, i.e. the collision energy for perfectly collimated, monoenergetic beams. The convolution of a simulated $E_{CM}^{-1}$ cross section with the angular and energetic distribution of the beams (blue lines) follows the experimental data at low collision energies.

In conclusion, we have presented the first measurements of absolute \textit{state-selective} mutual neutralization cross sections \textit{at subthermal energies}. This has been demonstrated for O$^-$ + N$^+$ and O$^-$ + O$^+$ collisions at energies ranging from 5 meV to 2 eV. The measured total mutual neutralization cross sections above 1 eV are in very good agreement with previous measurements from Hayton and Peart \cite{Hayton}. We calculated the cross sections and branching ratios down to 1~meV using the method by Zhou and Dickinson \cite{Zhou1997} to arrive at coupling strengths, which were then used in multi-channel LZ calculations. Using this procedure, we found relatively good agreement for O$^+$ + O$^-$, but in the case of N$^+$ + O$^-$ the population of the channel labeled 2s 2p$^4$ $^4$P could only be accounted for if the strong configuration interaction in this channel was considered. We expect that results at subthermal energies of the present quality will be crucial for modeling the charge balance and ion excitation energies in cold astrophysical environments.

In the present experiment, we could see small contributions from metastable excited states in the incoming ion beams. Such effects will almost always be a problem in merged beams experiments with \textit{molecular} ions. At the DESIREE (Double Electrostatic Storage Ring Experiment) infrastructure at Stockholm University \cite{Thomas,Schmidt2013}, beams of anions and cations can be stored for long times, relax to the lowest quantum states \cite{Schmidt2017} and be merged for studies of mutual neutralization at very low temperatures and well defined collision energies. Our present studies with atomic reactants will serve to benchmark similar experiments conducted with DESIREE. 

The authors thank M. Godefroid and J. Loreau for advice concerning the MCHF and Landau-Zener calculations and \AA{}. Larson for stimulating discussions. This work was supported by the Fonds de la Recherche Scientifique--FNRS (IISN Contract No. 4.4504.10) and by the Swedish Research Council (Contract No. 2017-00621, 621-2014-4501 and 621-2015-04990). XU is Senior Research Associate of the Fonds de la Recherche Scientifique--FNRS. TL is funded by a fellowship of the Fonds pour la Formation à la Recherche dans l'Industrie et dans l'Agriculture-FRIA.

\bibliography{mnCON}

\begin{thebibliography}{43}%
\makeatletter
\providecommand \@ifxundefined [1]{%
 \@ifx{#1\undefined}
}%
\providecommand \@ifnum [1]{%
 \ifnum #1\expandafter \@firstoftwo
 \else \expandafter \@secondoftwo
 \fi
}%
\providecommand \@ifx [1]{%
 \ifx #1\expandafter \@firstoftwo
 \else \expandafter \@secondoftwo
 \fi
}%
\providecommand \natexlab [1]{#1}%
\providecommand \enquote  [1]{``#1''}%
\providecommand \bibnamefont  [1]{#1}%
\providecommand \bibfnamefont [1]{#1}%
\providecommand \citenamefont [1]{#1}%
\providecommand \href@noop [0]{\@secondoftwo}%
\providecommand \href [0]{\begingroup \@sanitize@url \@href}%
\providecommand \@href[1]{\@@startlink{#1}\@@href}%
\providecommand \@@href[1]{\endgroup#1\@@endlink}%
\providecommand \@sanitize@url [0]{\catcode `\\12\catcode `\$12\catcode
  `\&12\catcode `\#12\catcode `\^12\catcode `\_12\catcode `\%12\relax}%
\providecommand \@@startlink[1]{}%
\providecommand \@@endlink[0]{}%
\providecommand \url  [0]{\begingroup\@sanitize@url \@url }%
\providecommand \@url [1]{\endgroup\@href {#1}{\urlprefix }}%
\providecommand \urlprefix  [0]{URL }%
\providecommand \Eprint [0]{\href }%
\providecommand \doibase [0]{http://dx.doi.org/}%
\providecommand \selectlanguage [0]{\@gobble}%
\providecommand \bibinfo  [0]{\@secondoftwo}%
\providecommand \bibfield  [0]{\@secondoftwo}%
\providecommand \translation [1]{[#1]}%
\providecommand \BibitemOpen [0]{}%
\providecommand \bibitemStop [0]{}%
\providecommand \bibitemNoStop [0]{.\EOS\space}%
\providecommand \EOS [0]{\spacefactor3000\relax}%
\providecommand \BibitemShut  [1]{\csname bibitem#1\endcsname}%
\let\auto@bib@innerbib\@empty
\bibitem [{\citenamefont {{Smith}}\ and\ \citenamefont
  {{Spanel}}(1995)}]{Smit95}%
  \BibitemOpen
  \bibfield  {author} {\bibinfo {author} {\bibfnamefont {D.}~\bibnamefont
  {{Smith}}}\ and\ \bibinfo {author} {\bibfnamefont {P.}~\bibnamefont
  {{Spanel}}},\ }\href {\doibase 10.1002/mas.1280140403} {\bibfield  {journal}
  {\bibinfo  {journal} {Mass Spectrom. Rev.}\ }\textbf {\bibinfo {volume}
  {14}},\ \bibinfo {pages} {255} (\bibinfo {year} {1995})}\BibitemShut
  {NoStop}%
\bibitem [{\citenamefont {Chutjian}\ \emph {et~al.}(1996)\citenamefont
  {Chutjian}, \citenamefont {Garscadden},\ and\ \citenamefont
  {Wadehra}}]{Chutjian}%
  \BibitemOpen
  \bibfield  {author} {\bibinfo {author} {\bibfnamefont {A.}~\bibnamefont
  {Chutjian}}, \bibinfo {author} {\bibfnamefont {A.}~\bibnamefont
  {Garscadden}}, \ and\ \bibinfo {author} {\bibfnamefont {J.}~\bibnamefont
  {Wadehra}},\ }\href {\doibase 10.1016/0370-1573(95)00022-4} {\bibfield
  {journal} {\bibinfo  {journal} {Phys. Reports}\ }\textbf {\bibinfo {volume}
  {264}},\ \bibinfo {pages} {393–470} (\bibinfo {year} {1996})}\BibitemShut
  {NoStop}%
\bibitem [{\citenamefont {{Larsson}}\ \emph {et~al.}(2012)\citenamefont
  {{Larsson}}, \citenamefont {{Geppert}},\ and\ \citenamefont
  {{Nyman}}}]{Larsson}%
  \BibitemOpen
  \bibfield  {author} {\bibinfo {author} {\bibfnamefont {M.}~\bibnamefont
  {{Larsson}}}, \bibinfo {author} {\bibfnamefont {W.~D.}\ \bibnamefont
  {{Geppert}}}, \ and\ \bibinfo {author} {\bibfnamefont {G.}~\bibnamefont
  {{Nyman}}},\ }\href {\doibase 10.1088/0034-4885/75/6/066901} {\bibfield
  {journal} {\bibinfo  {journal} {Rep. Prog. Phys.}\ }\textbf {\bibinfo
  {volume} {75}},\ \bibinfo {pages} {066901} (\bibinfo {year}
  {2012})}\BibitemShut {NoStop}%
\bibitem [{\citenamefont {Vuitton}\ \emph {et~al.}(2009)\citenamefont
  {Vuitton}, \citenamefont {Lavvas}, \citenamefont {Yelle}, \citenamefont
  {Galand}, \citenamefont {Wellbrock}, \citenamefont {Lewis}, \citenamefont
  {Coates},\ and\ \citenamefont {Wahlund}}]{Vuitton}%
  \BibitemOpen
  \bibfield  {author} {\bibinfo {author} {\bibfnamefont {V.}~\bibnamefont
  {Vuitton}}, \bibinfo {author} {\bibfnamefont {P.}~\bibnamefont {Lavvas}},
  \bibinfo {author} {\bibfnamefont {R.~V.}\ \bibnamefont {Yelle}}, \bibinfo
  {author} {\bibfnamefont {M.}~\bibnamefont {Galand}}, \bibinfo {author}
  {\bibfnamefont {A.}~\bibnamefont {Wellbrock}}, \bibinfo {author}
  {\bibfnamefont {G.~R.}\ \bibnamefont {Lewis}}, \bibinfo {author}
  {\bibfnamefont {A.~J.}\ \bibnamefont {Coates}}, \ and\ \bibinfo {author}
  {\bibfnamefont {J.~E.}\ \bibnamefont {Wahlund}},\ }\href {\doibase
  10.1016/j.pss.2009.04.004} {\bibfield  {journal} {\bibinfo  {journal}
  {Planet. Space Sci.}\ }\textbf {\bibinfo {volume} {57}},\ \bibinfo {pages}
  {1558} (\bibinfo {year} {2009})}\BibitemShut {NoStop}%
\bibitem [{\citenamefont {{Wildt}}(1939)}]{Wildt}%
  \BibitemOpen
  \bibfield  {author} {\bibinfo {author} {\bibfnamefont {R.}~\bibnamefont
  {{Wildt}}},\ }\href {\doibase 10.1086/144048} {\bibfield  {journal} {\bibinfo
   {journal} {Astrophys. J.}\ }\textbf {\bibinfo {volume} {89}},\ \bibinfo
  {pages} {295} (\bibinfo {year} {1939})}\BibitemShut {NoStop}%
\bibitem [{\citenamefont {{Rau}}(1996)}]{Rau}%
  \BibitemOpen
  \bibfield  {author} {\bibinfo {author} {\bibfnamefont {A.~R.~P.}\
  \bibnamefont {{Rau}}},\ }\href {\doibase 10.1007/BF02702300} {\bibfield
  {journal} {\bibinfo  {journal} {J. Astrophys. Astr.}\ }\textbf {\bibinfo
  {volume} {17}},\ \bibinfo {pages} {113} (\bibinfo {year} {1996})}\BibitemShut
  {NoStop}%
\bibitem [{\citenamefont {Chandrasekhar}\ and\ \citenamefont
  {Breen}(1946)}]{Chandrasekhar}%
  \BibitemOpen
  \bibfield  {author} {\bibinfo {author} {\bibfnamefont {S.}~\bibnamefont
  {Chandrasekhar}}\ and\ \bibinfo {author} {\bibfnamefont {F.~H.}\ \bibnamefont
  {Breen}},\ }\href {\doibase 10.1086/144874} {\bibfield  {journal} {\bibinfo
  {journal} {Astrophys. J.}\ }\textbf {\bibinfo {volume} {104}},\ \bibinfo
  {pages} {430} (\bibinfo {year} {1946})}\BibitemShut {NoStop}%
\bibitem [{\citenamefont {{Massey}}\ and\ \citenamefont
  {{Bates}}(1940)}]{Massey40}%
  \BibitemOpen
  \bibfield  {author} {\bibinfo {author} {\bibfnamefont {H.~S.~W.}\
  \bibnamefont {{Massey}}}\ and\ \bibinfo {author} {\bibfnamefont {D.~R.}\
  \bibnamefont {{Bates}}},\ }\href {\doibase 10.1086/144157} {\bibfield
  {journal} {\bibinfo  {journal} {Astrophys. J.}\ }\textbf {\bibinfo {volume}
  {91}},\ \bibinfo {pages} {202} (\bibinfo {year} {1940})}\BibitemShut
  {NoStop}%
\bibitem [{\citenamefont {Massey}(1950)}]{Massey50}%
  \BibitemOpen
  \bibfield  {author} {\bibinfo {author} {\bibfnamefont {H.}~\bibnamefont
  {Massey}},\ }\href@noop {} {\emph {\bibinfo {title} {Negative Ions}}},\
  \bibinfo {edition} {2nd}\ ed.\ (\bibinfo  {publisher} {Cambridge University
  Press},\ \bibinfo {year} {1950})\BibitemShut {NoStop}%
\bibitem [{\citenamefont {{Branscomb}}\ and\ \citenamefont
  {{Pagel}}(1958)}]{Branscomb}%
  \BibitemOpen
  \bibfield  {author} {\bibinfo {author} {\bibfnamefont {L.~M.}\ \bibnamefont
  {{Branscomb}}}\ and\ \bibinfo {author} {\bibfnamefont {B.~E.~J.}\
  \bibnamefont {{Pagel}}},\ }\href {\doibase 10.1093/mnras/118.3.258}
  {\bibfield  {journal} {\bibinfo  {journal} {Mon. Not. R. Astron. Soc.}\
  }\textbf {\bibinfo {volume} {118}},\ \bibinfo {pages} {258} (\bibinfo {year}
  {1958})}\BibitemShut {NoStop}%
\bibitem [{\citenamefont {{Vardya}}(1967)}]{Vardya}%
  \BibitemOpen
  \bibfield  {author} {\bibinfo {author} {\bibfnamefont {M.~S.}\ \bibnamefont
  {{Vardya}}},\ }\href {http://adsabs.harvard.edu/abs/1967MmRAS..71..249V}
  {\bibfield  {journal} {\bibinfo  {journal} {Mem. R. astr. Soc.}\ }\textbf
  {\bibinfo {volume} {71}},\ \bibinfo {pages} {249} (\bibinfo {year}
  {1967})}\BibitemShut {NoStop}%
\bibitem [{\citenamefont {{Dalgarno}}\ and\ \citenamefont
  {{McCray}}(1973)}]{DalgMcCray}%
  \BibitemOpen
  \bibfield  {author} {\bibinfo {author} {\bibfnamefont {A.}~\bibnamefont
  {{Dalgarno}}}\ and\ \bibinfo {author} {\bibfnamefont {R.~A.}\ \bibnamefont
  {{McCray}}},\ }\href {\doibase 10.1086/152032} {\bibfield  {journal}
  {\bibinfo  {journal} {Astrophys. J.}\ }\textbf {\bibinfo {volume} {181}},\
  \bibinfo {pages} {95} (\bibinfo {year} {1973})}\BibitemShut {NoStop}%
\bibitem [{\citenamefont {{Herbst}}(1981)}]{Herbst81}%
  \BibitemOpen
  \bibfield  {author} {\bibinfo {author} {\bibfnamefont {E.}~\bibnamefont
  {{Herbst}}},\ }\href {\doibase 10.1038/289656a0} {\bibfield  {journal}
  {\bibinfo  {journal} {Nature}\ }\textbf {\bibinfo {volume} {289}},\ \bibinfo
  {pages} {656} (\bibinfo {year} {1981})}\BibitemShut {NoStop}%
\bibitem [{\citenamefont {{McCarthy}}\ \emph {et~al.}(2006)\citenamefont
  {{McCarthy}}, \citenamefont {{Gottlieb}}, \citenamefont {{Gupta}},\ and\
  \citenamefont {{Thaddeus}}}]{McCarthy}%
  \BibitemOpen
  \bibfield  {author} {\bibinfo {author} {\bibfnamefont {M.~C.}\ \bibnamefont
  {{McCarthy}}}, \bibinfo {author} {\bibfnamefont {C.~A.}\ \bibnamefont
  {{Gottlieb}}}, \bibinfo {author} {\bibfnamefont {H.}~\bibnamefont {{Gupta}}},
  \ and\ \bibinfo {author} {\bibfnamefont {P.}~\bibnamefont {{Thaddeus}}},\
  }\href {\doibase 10.1086/510238} {\bibfield  {journal} {\bibinfo  {journal}
  {Astrophys. J.}\ }\textbf {\bibinfo {volume} {652}},\ \bibinfo {pages} {L141}
  (\bibinfo {year} {2006})}\BibitemShut {NoStop}%
\bibitem [{\citenamefont {{Cernicharo}}\ \emph {et~al.}(2007)\citenamefont
  {{Cernicharo}}, \citenamefont {{Gu\'elin}}, \citenamefont {{Agundez}},
  \citenamefont {{Kawaguchi}}, \citenamefont {{McCarthy}},\ and\ \citenamefont
  {{Thaddeus}}}]{Cernicharo2007}%
  \BibitemOpen
  \bibfield  {author} {\bibinfo {author} {\bibfnamefont {J.}~\bibnamefont
  {{Cernicharo}}}, \bibinfo {author} {\bibfnamefont {M.}~\bibnamefont
  {{Gu\'elin}}}, \bibinfo {author} {\bibfnamefont {M.}~\bibnamefont
  {{Agundez}}}, \bibinfo {author} {\bibfnamefont {K.}~\bibnamefont
  {{Kawaguchi}}}, \bibinfo {author} {\bibfnamefont {M.}~\bibnamefont
  {{McCarthy}}}, \ and\ \bibinfo {author} {\bibfnamefont {P.}~\bibnamefont
  {{Thaddeus}}},\ }\href {\doibase 10.1051/0004-6361:20077415} {\bibfield
  {journal} {\bibinfo  {journal} {Astron. \& Astrophys.}\ }\textbf {\bibinfo
  {volume} {467}},\ \bibinfo {pages} {L37} (\bibinfo {year}
  {2007})}\BibitemShut {NoStop}%
\bibitem [{\citenamefont {{Br\"unken}}\ \emph {et~al.}(2007)\citenamefont
  {{Br\"unken}}, \citenamefont {{Gupta}}, \citenamefont {{Gottlieb}},
  \citenamefont {{McCarthy}},\ and\ \citenamefont {{Thaddeus}}}]{Brunken}%
  \BibitemOpen
  \bibfield  {author} {\bibinfo {author} {\bibfnamefont {S.}~\bibnamefont
  {{Br\"unken}}}, \bibinfo {author} {\bibfnamefont {H.}~\bibnamefont
  {{Gupta}}}, \bibinfo {author} {\bibfnamefont {C.~A.}\ \bibnamefont
  {{Gottlieb}}}, \bibinfo {author} {\bibfnamefont {M.}~\bibnamefont
  {{McCarthy}}}, \ and\ \bibinfo {author} {\bibfnamefont {P.}~\bibnamefont
  {{Thaddeus}}},\ }\href {\doibase 10.1086/520703} {\bibfield  {journal}
  {\bibinfo  {journal} {Astrophys. J.}\ }\textbf {\bibinfo {volume} {664}},\
  \bibinfo {pages} {L43} (\bibinfo {year} {2007})}\BibitemShut {NoStop}%
\bibitem [{\citenamefont {{Thaddeus}}\ \emph {et~al.}(2008)\citenamefont
  {{Thaddeus}}, \citenamefont {{Gottlieb}}, \citenamefont {{Gupta}},
  \citenamefont {{Br\"unken}}, \citenamefont {{McCarthy}}, \citenamefont
  {{Ag\'undez}}, \citenamefont {{Gu\'elin}},\ and\ \citenamefont
  {{Cernicharo}}}]{Thaddeus}%
  \BibitemOpen
  \bibfield  {author} {\bibinfo {author} {\bibfnamefont {P.}~\bibnamefont
  {{Thaddeus}}}, \bibinfo {author} {\bibfnamefont {C.~A.}\ \bibnamefont
  {{Gottlieb}}}, \bibinfo {author} {\bibfnamefont {H.}~\bibnamefont {{Gupta}}},
  \bibinfo {author} {\bibfnamefont {S.}~\bibnamefont {{Br\"unken}}}, \bibinfo
  {author} {\bibfnamefont {M.~C.}\ \bibnamefont {{McCarthy}}}, \bibinfo
  {author} {\bibfnamefont {M.}~\bibnamefont {{Ag\'undez}}}, \bibinfo {author}
  {\bibfnamefont {M.}~\bibnamefont {{Gu\'elin}}}, \ and\ \bibinfo {author}
  {\bibfnamefont {J.}~\bibnamefont {{Cernicharo}}},\ }\href {\doibase
  10.1086/528947} {\bibfield  {journal} {\bibinfo  {journal} {Astrophys. J.}\
  }\textbf {\bibinfo {volume} {677}},\ \bibinfo {pages} {1132} (\bibinfo {year}
  {2008})}\BibitemShut {NoStop}%
\bibitem [{\citenamefont {{Cernicharo}}\ \emph {et~al.}(2008)\citenamefont
  {{Cernicharo}}, \citenamefont {{Gu\'elin}}, \citenamefont {{Agundez}},
  \citenamefont {{McCarthy}},\ and\ \citenamefont
  {{Thaddeus}}}]{Cernicharo2008}%
  \BibitemOpen
  \bibfield  {author} {\bibinfo {author} {\bibfnamefont {J.}~\bibnamefont
  {{Cernicharo}}}, \bibinfo {author} {\bibfnamefont {M.}~\bibnamefont
  {{Gu\'elin}}}, \bibinfo {author} {\bibfnamefont {M.}~\bibnamefont
  {{Agundez}}}, \bibinfo {author} {\bibfnamefont {M.}~\bibnamefont
  {{McCarthy}}}, \ and\ \bibinfo {author} {\bibfnamefont {P.}~\bibnamefont
  {{Thaddeus}}},\ }\href {\doibase 10.1086/595583} {\bibfield  {journal}
  {\bibinfo  {journal} {Astrophys. J. Lett.}\ }\textbf {\bibinfo {volume}
  {688}},\ \bibinfo {pages} {L83} (\bibinfo {year} {2008})}\BibitemShut
  {NoStop}%
\bibitem [{\citenamefont {{Agundez}}\ \emph {et~al.}(2010)\citenamefont
  {{Agundez}}, \citenamefont {{Cernicharo}}, \citenamefont {{Gu\'elin}},
  \citenamefont {{Kahane}}, \citenamefont {{Roueff}}, \citenamefont {{Klos}},
  \citenamefont {{Aoiz}}, \citenamefont {{Lique}}, \citenamefont {{Marcelino}},
  \citenamefont {{Goicoechea}} \emph {et~al.}}]{Agundez}%
  \BibitemOpen
  \bibfield  {author} {\bibinfo {author} {\bibfnamefont {M.}~\bibnamefont
  {{Agundez}}}, \bibinfo {author} {\bibfnamefont {J.}~\bibnamefont
  {{Cernicharo}}}, \bibinfo {author} {\bibfnamefont {M.}~\bibnamefont
  {{Gu\'elin}}}, \bibinfo {author} {\bibfnamefont {C.}~\bibnamefont
  {{Kahane}}}, \bibinfo {author} {\bibfnamefont {E.}~\bibnamefont {{Roueff}}},
  \bibinfo {author} {\bibfnamefont {J.}~\bibnamefont {{Klos}}}, \bibinfo
  {author} {\bibfnamefont {F.~J.}\ \bibnamefont {{Aoiz}}}, \bibinfo {author}
  {\bibfnamefont {F.}~\bibnamefont {{Lique}}}, \bibinfo {author} {\bibfnamefont
  {N.}~\bibnamefont {{Marcelino}}}, \bibinfo {author} {\bibfnamefont {J.~R.}\
  \bibnamefont {{Goicoechea}}},  \emph {et~al.},\ }\href {\doibase
  10.1051/0004-6361/201015186} {\bibfield  {journal} {\bibinfo  {journal}
  {Astron. \& Astrophys.}\ }\textbf {\bibinfo {volume} {517}},\ \bibinfo
  {pages} {L2} (\bibinfo {year} {2010})}\BibitemShut {NoStop}%
\bibitem [{\citenamefont {{Chaizy}}\ \emph {et~al.}(1991)\citenamefont
  {{Chaizy}}, \citenamefont {{R\`eme}}, \citenamefont {{Sauvaud}},
  \citenamefont {{d'Uston}}, \citenamefont {{Lin}}, \citenamefont {{Larson}},
  \citenamefont {{Mitchell}}, \citenamefont {{Anderson}}, \citenamefont
  {{Carlson}}, \citenamefont {{Korth}},\ and\ \citenamefont
  {{Mendis}}}]{Chaizy}%
  \BibitemOpen
  \bibfield  {author} {\bibinfo {author} {\bibfnamefont {P.}~\bibnamefont
  {{Chaizy}}}, \bibinfo {author} {\bibfnamefont {H.}~\bibnamefont {{R\`eme}}},
  \bibinfo {author} {\bibfnamefont {J.~A.}\ \bibnamefont {{Sauvaud}}}, \bibinfo
  {author} {\bibfnamefont {C.}~\bibnamefont {{d'Uston}}}, \bibinfo {author}
  {\bibfnamefont {R.~P.}\ \bibnamefont {{Lin}}}, \bibinfo {author}
  {\bibfnamefont {D.~E.}\ \bibnamefont {{Larson}}}, \bibinfo {author}
  {\bibfnamefont {D.~L.}\ \bibnamefont {{Mitchell}}}, \bibinfo {author}
  {\bibfnamefont {K.~A.}\ \bibnamefont {{Anderson}}}, \bibinfo {author}
  {\bibfnamefont {C.~W.}\ \bibnamefont {{Carlson}}}, \bibinfo {author}
  {\bibfnamefont {A.}~\bibnamefont {{Korth}}}, \ and\ \bibinfo {author}
  {\bibfnamefont {D.~A.}\ \bibnamefont {{Mendis}}},\ }\href {\doibase
  10.1038/349393a0} {\bibfield  {journal} {\bibinfo  {journal} {Nature}\
  }\textbf {\bibinfo {volume} {349}},\ \bibinfo {pages} {393} (\bibinfo {year}
  {1991})}\BibitemShut {NoStop}%
\bibitem [{\citenamefont {{Liu}}(1998)}]{liu1998}%
  \BibitemOpen
  \bibfield  {author} {\bibinfo {author} {\bibfnamefont {W.}~\bibnamefont
  {{Liu}}},\ }\href {\doibase 10.1086/305390} {\bibfield  {journal} {\bibinfo
  {journal} {Astrophys. J.}\ }\textbf {\bibinfo {volume} {496}},\ \bibinfo
  {pages} {967} (\bibinfo {year} {1998})}\BibitemShut {NoStop}%
\bibitem [{\citenamefont {{Harada}}\ and\ \citenamefont
  {{Herbst}}(2008)}]{Harada}%
  \BibitemOpen
  \bibfield  {author} {\bibinfo {author} {\bibfnamefont {N.}~\bibnamefont
  {{Harada}}}\ and\ \bibinfo {author} {\bibfnamefont {E.}~\bibnamefont
  {{Herbst}}},\ }\href {\doibase 10.1086/590468} {\bibfield  {journal}
  {\bibinfo  {journal} {Astrophys. J.}\ }\textbf {\bibinfo {volume} {685}},\
  \bibinfo {pages} {272} (\bibinfo {year} {2008})}\BibitemShut {NoStop}%
\bibitem [{\citenamefont {Flower}\ \emph {et~al.}(2007)\citenamefont {Flower},
  \citenamefont {Pineau Des~For\^ets},\ and\ \citenamefont
  {Walmsley}}]{Flower}%
  \BibitemOpen
  \bibfield  {author} {\bibinfo {author} {\bibfnamefont {D.~R.}\ \bibnamefont
  {Flower}}, \bibinfo {author} {\bibfnamefont {G.}~\bibnamefont {Pineau
  Des~For\^ets}}, \ and\ \bibinfo {author} {\bibfnamefont {C.~M.}\ \bibnamefont
  {Walmsley}},\ }\href {\doibase 10.1051/0004-6361:20078138} {\bibfield
  {journal} {\bibinfo  {journal} {Astron. \& Astrophys.}\ }\textbf {\bibinfo
  {volume} {474}},\ \bibinfo {pages} {923} (\bibinfo {year}
  {2007})}\BibitemShut {NoStop}%
\bibitem [{\citenamefont {{Cordiner}}\ and\ \citenamefont
  {{Charnley}}(2012)}]{Cordiner12}%
  \BibitemOpen
  \bibfield  {author} {\bibinfo {author} {\bibfnamefont {M.~A.}\ \bibnamefont
  {{Cordiner}}}\ and\ \bibinfo {author} {\bibfnamefont {S.~B.}\ \bibnamefont
  {{Charnley}}},\ }\href {\doibase 10.1088/0004-637X/749/2/120} {\bibfield
  {journal} {\bibinfo  {journal} {Astrophys. J.}\ }\textbf {\bibinfo {volume}
  {749}},\ \bibinfo {pages} {120} (\bibinfo {year} {2012})}\BibitemShut
  {NoStop}%
\bibitem [{\citenamefont {Millar}\ \emph {et~al.}(2007)\citenamefont {Millar},
  \citenamefont {Walsh}, \citenamefont {Cordiner}, \citenamefont
  {N\'i~Chuim\'in},\ and\ \citenamefont {Herbst}}]{Millar}%
  \BibitemOpen
  \bibfield  {author} {\bibinfo {author} {\bibfnamefont {T.}~\bibnamefont
  {Millar}}, \bibinfo {author} {\bibfnamefont {C.}~\bibnamefont {Walsh}},
  \bibinfo {author} {\bibfnamefont {M.}~\bibnamefont {Cordiner}}, \bibinfo
  {author} {\bibfnamefont {R.}~\bibnamefont {N\'i~Chuim\'in}}, \ and\ \bibinfo
  {author} {\bibfnamefont {E.}~\bibnamefont {Herbst}},\ }\href {\doibase
  10.1086/519376} {\bibfield  {journal} {\bibinfo  {journal} {Astrophys. J.}\
  }\textbf {\bibinfo {volume} {662}},\ \bibinfo {pages} {L87} (\bibinfo {year}
  {2007})}\BibitemShut {NoStop}%
\bibitem [{\citenamefont {{Wakelam}}\ and\ \citenamefont
  {{Herbst}}(2008)}]{Wake08}%
  \BibitemOpen
  \bibfield  {author} {\bibinfo {author} {\bibfnamefont {V.}~\bibnamefont
  {{Wakelam}}}\ and\ \bibinfo {author} {\bibfnamefont {E.}~\bibnamefont
  {{Herbst}}},\ }\href {\doibase 10.1086/587734} {\bibfield  {journal}
  {\bibinfo  {journal} {Astrophys. J.}\ }\textbf {\bibinfo {volume} {680}},\
  \bibinfo {pages} {371} (\bibinfo {year} {2008})}\BibitemShut {NoStop}%
\bibitem [{\citenamefont {{Walsh}}\ \emph {et~al.}(2009)\citenamefont
  {{Walsh}}, \citenamefont {{Harada}}, \citenamefont {{Herbst}},\ and\
  \citenamefont {{Millar}}}]{Walsh}%
  \BibitemOpen
  \bibfield  {author} {\bibinfo {author} {\bibfnamefont {C.}~\bibnamefont
  {{Walsh}}}, \bibinfo {author} {\bibfnamefont {N.}~\bibnamefont {{Harada}}},
  \bibinfo {author} {\bibfnamefont {E.}~\bibnamefont {{Herbst}}}, \ and\
  \bibinfo {author} {\bibfnamefont {T.~J.}\ \bibnamefont {{Millar}}},\ }\href
  {\doibase 10.1088/0004-637X/700/1/752} {\bibfield  {journal} {\bibinfo
  {journal} {Astrophys. J.}\ }\textbf {\bibinfo {volume} {700}},\ \bibinfo
  {pages} {725} (\bibinfo {year} {2009})}\BibitemShut {NoStop}%
\bibitem [{\citenamefont {{Wakelam}}\ \emph {et~al.}(2012)\citenamefont
  {{Wakelam}}, \citenamefont {{Herbst}}, \citenamefont {{Loison}},
  \citenamefont {{Smith}}, \citenamefont {{Chandrasekaran}}, \citenamefont
  {{Pavone}}, \citenamefont {{Adams}}, \citenamefont {{Bacchus-Montabonel}},
  \citenamefont {{Bergeat}}, \citenamefont {{B{\'e}roff}}, \citenamefont
  {{Bierbaum}} \emph {et~al.}}]{Wake12a}%
  \BibitemOpen
  \bibfield  {author} {\bibinfo {author} {\bibfnamefont {V.}~\bibnamefont
  {{Wakelam}}}, \bibinfo {author} {\bibfnamefont {E.}~\bibnamefont {{Herbst}}},
  \bibinfo {author} {\bibfnamefont {J.-C.}\ \bibnamefont {{Loison}}}, \bibinfo
  {author} {\bibfnamefont {I.~W.~M.}\ \bibnamefont {{Smith}}}, \bibinfo
  {author} {\bibfnamefont {V.}~\bibnamefont {{Chandrasekaran}}}, \bibinfo
  {author} {\bibfnamefont {B.}~\bibnamefont {{Pavone}}}, \bibinfo {author}
  {\bibfnamefont {N.~G.}\ \bibnamefont {{Adams}}}, \bibinfo {author}
  {\bibfnamefont {M.-C.}\ \bibnamefont {{Bacchus-Montabonel}}}, \bibinfo
  {author} {\bibfnamefont {A.}~\bibnamefont {{Bergeat}}}, \bibinfo {author}
  {\bibfnamefont {K.}~\bibnamefont {{B{\'e}roff}}}, \bibinfo {author}
  {\bibfnamefont {V.~M.}\ \bibnamefont {{Bierbaum}}},  \emph {et~al.},\ }\href
  {\doibase 10.1088/0067-0049/199/1/21} {\bibfield  {journal} {\bibinfo
  {journal} {Astrophys. J. Suppl. Ser.}\ }\textbf {\bibinfo {volume} {199}},\
  \bibinfo {eid} {21} (\bibinfo {year} {2012})}\BibitemShut {NoStop}%
\bibitem [{\citenamefont {{McElroy}}\ \emph {et~al.}(2013)\citenamefont
  {{McElroy}}, \citenamefont {{Walsh}}, \citenamefont {{Markwick}},
  \citenamefont {{Cordiner}}, \citenamefont {{Smith}},\ and\ \citenamefont
  {{Millar}}}]{UMIST13}%
  \BibitemOpen
  \bibfield  {author} {\bibinfo {author} {\bibfnamefont {D.}~\bibnamefont
  {{McElroy}}}, \bibinfo {author} {\bibfnamefont {C.}~\bibnamefont {{Walsh}}},
  \bibinfo {author} {\bibfnamefont {A.~J.}\ \bibnamefont {{Markwick}}},
  \bibinfo {author} {\bibfnamefont {M.~A.}\ \bibnamefont {{Cordiner}}},
  \bibinfo {author} {\bibfnamefont {K.}~\bibnamefont {{Smith}}}, \ and\
  \bibinfo {author} {\bibfnamefont {T.~J.}\ \bibnamefont {{Millar}}},\ }\href
  {\doibase 10.1051/0004-6361/201220465} {\bibfield  {journal} {\bibinfo
  {journal} {Astron. \& Astrophys.}\ }\textbf {\bibinfo {volume} {550}},\
  \bibinfo {eid} {A36} (\bibinfo {year} {2013})},\ \Eprint
  {http://arxiv.org/abs/1212.6362} {arXiv:1212.6362 [astro-ph.SR]} \BibitemShut
  {NoStop}%
\bibitem [{\citenamefont {{Cordiner}}\ and\ \citenamefont
  {{Charnley}}(2014)}]{Cordiner}%
  \BibitemOpen
  \bibfield  {author} {\bibinfo {author} {\bibfnamefont {M.~A.}\ \bibnamefont
  {{Cordiner}}}\ and\ \bibinfo {author} {\bibfnamefont {S.~B.}\ \bibnamefont
  {{Charnley}}},\ }\href {\doibase 10.1111/maps.12082} {\bibfield  {journal}
  {\bibinfo  {journal} {Meteorit. \& Planet. Sci.}\ }\textbf {\bibinfo {volume}
  {49}},\ \bibinfo {pages} {21} (\bibinfo {year} {2014})}\BibitemShut {NoStop}%
\bibitem [{\citenamefont {{Hayton}}\ and\ \citenamefont
  {{Peart}}(1993)}]{Hayton}%
  \BibitemOpen
  \bibfield  {author} {\bibinfo {author} {\bibfnamefont {D.~A.}\ \bibnamefont
  {{Hayton}}}\ and\ \bibinfo {author} {\bibfnamefont {B.}~\bibnamefont
  {{Peart}}},\ }\href {\doibase 10.1088/0953-4075/26/17/020} {\bibfield
  {journal} {\bibinfo  {journal} {J. Phys. B}\ }\textbf {\bibinfo {volume}
  {26}},\ \bibinfo {pages} {2879} (\bibinfo {year} {1993})}\BibitemShut
  {NoStop}%
\bibitem [{\citenamefont {{Terao}}\ \emph {et~al.}(1986)\citenamefont
  {{Terao}}, \citenamefont {{Sz{\"u}cs}}, \citenamefont {{Cherkani}},
  \citenamefont {{Brouillard}}, \citenamefont {{Allan}}, \citenamefont
  {{Harel}},\ and\ \citenamefont {{Salin}}}]{Terao}%
  \BibitemOpen
  \bibfield  {author} {\bibinfo {author} {\bibfnamefont {M.}~\bibnamefont
  {{Terao}}}, \bibinfo {author} {\bibfnamefont {S.}~\bibnamefont
  {{Sz{\"u}cs}}}, \bibinfo {author} {\bibfnamefont {M.}~\bibnamefont
  {{Cherkani}}}, \bibinfo {author} {\bibfnamefont {F.}~\bibnamefont
  {{Brouillard}}}, \bibinfo {author} {\bibfnamefont {R.~J.}\ \bibnamefont
  {{Allan}}}, \bibinfo {author} {\bibfnamefont {C.}~\bibnamefont {{Harel}}}, \
  and\ \bibinfo {author} {\bibfnamefont {A.}~\bibnamefont {{Salin}}},\ }\href
  {\doibase doi:10.1209/0295-5075/1/3/005} {\bibfield  {journal} {\bibinfo
  {journal} {Europhys. Lett.}\ }\textbf {\bibinfo {volume} {1}},\ \bibinfo
  {pages} {123} (\bibinfo {year} {1986})}\BibitemShut {NoStop}%
\bibitem [{\citenamefont {Nkambule}\ \emph {et~al.}(2016)\citenamefont
  {Nkambule}, \citenamefont {Elander}, \citenamefont {Larson}, \citenamefont
  {Lecointre},\ and\ \citenamefont {Urbain}}]{Nkambule2016}%
  \BibitemOpen
  \bibfield  {author} {\bibinfo {author} {\bibfnamefont {S.~M.}\ \bibnamefont
  {Nkambule}}, \bibinfo {author} {\bibfnamefont {N.}~\bibnamefont {Elander}},
  \bibinfo {author} {\bibfnamefont {A.}~\bibnamefont {Larson}}, \bibinfo
  {author} {\bibfnamefont {J.}~\bibnamefont {Lecointre}}, \ and\ \bibinfo
  {author} {\bibfnamefont {X.}~\bibnamefont {Urbain}},\ }\href {\doibase
  10.1103/PhysRevA.93.032701} {\bibfield  {journal} {\bibinfo  {journal} {Phys.
  Rev. A}\ }\textbf {\bibinfo {volume} {93}},\ \bibinfo {pages} {032701}
  (\bibinfo {year} {2016})}\BibitemShut {NoStop}%
\bibitem [{\citenamefont {{Nzeyimana}}\ \emph {et~al.}(2002)\citenamefont
  {{Nzeyimana}}, \citenamefont {{Naji}}, \citenamefont {{Urbain}},\ and\
  \citenamefont {{Le Padellec}}}]{Nzeyi}%
  \BibitemOpen
  \bibfield  {author} {\bibinfo {author} {\bibfnamefont {T.}~\bibnamefont
  {{Nzeyimana}}}, \bibinfo {author} {\bibfnamefont {E.~A.}\ \bibnamefont
  {{Naji}}}, \bibinfo {author} {\bibfnamefont {X.}~\bibnamefont {{Urbain}}}, \
  and\ \bibinfo {author} {\bibfnamefont {A.}~\bibnamefont {{Le Padellec}}},\
  }\href {\doibase 10.1140/epjd/e20020090} {\bibfield  {journal} {\bibinfo
  {journal} {Eur. Phys. J. D}\ }\textbf {\bibinfo {volume} {19}},\ \bibinfo
  {pages} {315} (\bibinfo {year} {2002})}\BibitemShut {NoStop}%
\bibitem [{\citenamefont {Fabre}(2005)}]{fabre}%
  \BibitemOpen
  \bibfield  {author} {\bibinfo {author} {\bibfnamefont {B.}~\bibnamefont
  {Fabre}},\ }\emph {\bibinfo {title} {Paquets d'onde vibrationnels cr\'e\'es
  par ionisation de H$_2$ en champ laser intense}},\ \href@noop {} {Ph.D.
  thesis},\ \bibinfo  {school} {Universit\'e catholique de Louvain} (\bibinfo
  {year} {2005})\BibitemShut {NoStop}%
\bibitem [{\citenamefont {{Sz{\"u}cs}}\ \emph {et~al.}(1984)\citenamefont
  {{Sz{\"u}cs}}, \citenamefont {{Karemera}}, \citenamefont {{Terao}},\ and\
  \citenamefont {{Brouillard}}}]{Szucs}%
  \BibitemOpen
  \bibfield  {author} {\bibinfo {author} {\bibfnamefont {S.}~\bibnamefont
  {{Sz{\"u}cs}}}, \bibinfo {author} {\bibfnamefont {M.}~\bibnamefont
  {{Karemera}}}, \bibinfo {author} {\bibfnamefont {M.}~\bibnamefont {{Terao}}},
  \ and\ \bibinfo {author} {\bibfnamefont {F.}~\bibnamefont {{Brouillard}}},\
  }\href {http://stacks.iop.org/0022-3700/17/i=8/a=021} {\bibfield  {journal}
  {\bibinfo  {journal} {J. Phys. B}\ }\textbf {\bibinfo {volume} {17}},\
  \bibinfo {pages} {1613} (\bibinfo {year} {1984})}\BibitemShut {NoStop}%
\bibitem [{\citenamefont {{Brouillard}}\ and\ \citenamefont
  {{Claeys}}(1983)}]{Brou83a}%
  \BibitemOpen
  \bibfield  {author} {\bibinfo {author} {\bibfnamefont {F.}~\bibnamefont
  {{Brouillard}}}\ and\ \bibinfo {author} {\bibfnamefont {W.}~\bibnamefont
  {{Claeys}}},\ }in\ \href@noop {} {\emph {\bibinfo {booktitle} {Physics of
  Ion-Ion and Electron-Ion Collisions}}},\ \bibinfo {editor} {edited by\
  \bibinfo {editor} {\bibfnamefont {F.}~\bibnamefont {{Brouillard}}}\ and\
  \bibinfo {editor} {\bibfnamefont {W.}~\bibnamefont {{Claeys}}}}\ (\bibinfo
  {publisher} {Plenum Press},\ \bibinfo {address} {New York},\ \bibinfo {year}
  {1983})\ pp.\ \bibinfo {pages} {415--459}\BibitemShut {NoStop}%
\bibitem [{\citenamefont {Zhou}\ and\ \citenamefont
  {Dickinson}(1997)}]{Zhou1997}%
  \BibitemOpen
  \bibfield  {author} {\bibinfo {author} {\bibfnamefont {X.}~\bibnamefont
  {Zhou}}\ and\ \bibinfo {author} {\bibfnamefont {A.}~\bibnamefont
  {Dickinson}},\ }\href {\doibase 10.1016/S0168-583X(96)00922-6} {\bibfield
  {journal} {\bibinfo  {journal} {Nucl. Instrum. Methods B}\ }\textbf {\bibinfo
  {volume} {124}},\ \bibinfo {pages} {5} (\bibinfo {year} {1997})}\BibitemShut
  {NoStop}%
\bibitem [{\citenamefont {Chibisov}\ and\ \citenamefont
  {Janev}(1988)}]{Chibisov1988}%
  \BibitemOpen
  \bibfield  {author} {\bibinfo {author} {\bibfnamefont {M.}~\bibnamefont
  {Chibisov}}\ and\ \bibinfo {author} {\bibfnamefont {R.}~\bibnamefont
  {Janev}},\ }\href {http://dx.doi.org/10.1016/S0370-1573(98)90002-3}
  {\bibfield  {journal} {\bibinfo  {journal} {Phys. Rep.}\ }\textbf {\bibinfo
  {volume} {166}},\ \bibinfo {pages} {1} (\bibinfo {year} {1988})}\BibitemShut
  {NoStop}%
\bibitem [{\citenamefont {Mitrushchenkov}\ \emph {et~al.}(2017)\citenamefont
  {Mitrushchenkov}, \citenamefont {Guitou}, \citenamefont {Belyaev},
  \citenamefont {Yakovleva}, \citenamefont {Spielfiedel},\ and\ \citenamefont
  {Feautrier}}]{Mitrushchenkov}%
  \BibitemOpen
  \bibfield  {author} {\bibinfo {author} {\bibfnamefont {A.}~\bibnamefont
  {Mitrushchenkov}}, \bibinfo {author} {\bibfnamefont {M.}~\bibnamefont
  {Guitou}}, \bibinfo {author} {\bibfnamefont {A.~K.}\ \bibnamefont {Belyaev}},
  \bibinfo {author} {\bibfnamefont {S.~A.}\ \bibnamefont {Yakovleva}}, \bibinfo
  {author} {\bibfnamefont {A.}~\bibnamefont {Spielfiedel}}, \ and\ \bibinfo
  {author} {\bibfnamefont {N.}~\bibnamefont {Feautrier}},\ }\href {\doibase
  10.1063/1.4973457} {\bibfield  {journal} {\bibinfo  {journal} {J. Chem.
  Phys.}\ }\textbf {\bibinfo {volume} {146}},\ \bibinfo {pages} {014304}
  (\bibinfo {year} {2017})}\BibitemShut {NoStop}%
\bibitem [{\citenamefont {{Thomas}}\ \emph {et~al.}(2011)\citenamefont
  {{Thomas}}, \citenamefont {{Schmidt}}, \citenamefont {{Andler}},
  \citenamefont {{Bj\"orkhage}}, \citenamefont {{Blom}}, \citenamefont
  {{Br\"annholm}}, \citenamefont {{B\"ackstr\"om}}, \citenamefont {{Danared}},
  \citenamefont {{Das}}, \citenamefont {{Haag}} \emph {et~al.}}]{Thomas}%
  \BibitemOpen
  \bibfield  {author} {\bibinfo {author} {\bibfnamefont {R.~D.}\ \bibnamefont
  {{Thomas}}}, \bibinfo {author} {\bibfnamefont {H.~T.}\ \bibnamefont
  {{Schmidt}}}, \bibinfo {author} {\bibfnamefont {G.}~\bibnamefont {{Andler}}},
  \bibinfo {author} {\bibfnamefont {M.}~\bibnamefont {{Bj\"orkhage}}}, \bibinfo
  {author} {\bibfnamefont {M.}~\bibnamefont {{Blom}}}, \bibinfo {author}
  {\bibfnamefont {L.}~\bibnamefont {{Br\"annholm}}}, \bibinfo {author}
  {\bibfnamefont {E.}~\bibnamefont {{B\"ackstr\"om}}}, \bibinfo {author}
  {\bibfnamefont {H.}~\bibnamefont {{Danared}}}, \bibinfo {author}
  {\bibfnamefont {S.}~\bibnamefont {{Das}}}, \bibinfo {author} {\bibfnamefont
  {N.}~\bibnamefont {{Haag}}},  \emph {et~al.},\ }\href {\doibase
  10.1063/1.3602928} {\bibfield  {journal} {\bibinfo  {journal} {Rev. Sci.
  Instrum.}\ }\textbf {\bibinfo {volume} {82}},\ \bibinfo {pages} {065112}
  (\bibinfo {year} {2011})}\BibitemShut {NoStop}%
\bibitem [{\citenamefont {Schmidt}\ \emph {et~al.}(2013)\citenamefont
  {Schmidt}, \citenamefont {Thomas}, \citenamefont {Gatchell}, \citenamefont
  {Ros\'en}, \citenamefont {Reinhed}, \citenamefont {L\"ofgren}, \citenamefont
  {Br\"annholm}, \citenamefont {Blom}, \citenamefont {Bj\"orkhage},
  \citenamefont {B\"ackstr\"om} \emph {et~al.}}]{Schmidt2013}%
  \BibitemOpen
  \bibfield  {author} {\bibinfo {author} {\bibfnamefont {H.~T.}\ \bibnamefont
  {Schmidt}}, \bibinfo {author} {\bibfnamefont {R.~D.}\ \bibnamefont {Thomas}},
  \bibinfo {author} {\bibfnamefont {M.}~\bibnamefont {Gatchell}}, \bibinfo
  {author} {\bibfnamefont {S.}~\bibnamefont {Ros\'en}}, \bibinfo {author}
  {\bibfnamefont {P.}~\bibnamefont {Reinhed}}, \bibinfo {author} {\bibfnamefont
  {P.}~\bibnamefont {L\"ofgren}}, \bibinfo {author} {\bibfnamefont
  {L.}~\bibnamefont {Br\"annholm}}, \bibinfo {author} {\bibfnamefont
  {M.}~\bibnamefont {Blom}}, \bibinfo {author} {\bibfnamefont {M.}~\bibnamefont
  {Bj\"orkhage}}, \bibinfo {author} {\bibfnamefont {E.}~\bibnamefont
  {B\"ackstr\"om}},  \emph {et~al.},\ }\href {\doibase 10.1063/1.4807702}
  {\bibfield  {journal} {\bibinfo  {journal} {Rev. Sci. Instrum.}\ }\textbf
  {\bibinfo {volume} {84}},\ \bibinfo {pages} {055115} (\bibinfo {year}
  {2013})}\BibitemShut {NoStop}%
\bibitem [{\citenamefont {Schmidt}\ \emph {et~al.}(2017)\citenamefont
  {Schmidt}, \citenamefont {Eklund}, \citenamefont {Chartkunchand},
  \citenamefont {Anderson}, \citenamefont {Kami\ifmmode~\acute{n}\else
  \'{n}\fi{}ska}, \citenamefont {de~Ruette}, \citenamefont {Thomas},
  \citenamefont {Kristiansson}, \citenamefont {Gatchell}, \citenamefont
  {Reinhed}, \citenamefont {Ros\'en}, \citenamefont {Simonsson}, \citenamefont
  {K\"allberg}, \citenamefont {L\"ofgren}, \citenamefont {Mannervik},
  \citenamefont {Zettergren},\ and\ \citenamefont {Cederquist}}]{Schmidt2017}%
  \BibitemOpen
  \bibfield  {author} {\bibinfo {author} {\bibfnamefont {H.~T.}\ \bibnamefont
  {Schmidt}}, \bibinfo {author} {\bibfnamefont {G.}~\bibnamefont {Eklund}},
  \bibinfo {author} {\bibfnamefont {K.~C.}\ \bibnamefont {Chartkunchand}},
  \bibinfo {author} {\bibfnamefont {E.~K.}\ \bibnamefont {Anderson}}, \bibinfo
  {author} {\bibfnamefont {M.}~\bibnamefont {Kami\ifmmode~\acute{n}\else
  \'{n}\fi{}ska}}, \bibinfo {author} {\bibfnamefont {N.}~\bibnamefont
  {de~Ruette}}, \bibinfo {author} {\bibfnamefont {R.~D.}\ \bibnamefont
  {Thomas}}, \bibinfo {author} {\bibfnamefont {M.~K.}\ \bibnamefont
  {Kristiansson}}, \bibinfo {author} {\bibfnamefont {M.}~\bibnamefont
  {Gatchell}}, \bibinfo {author} {\bibfnamefont {P.}~\bibnamefont {Reinhed}},
  \bibinfo {author} {\bibfnamefont {S.}~\bibnamefont {Ros\'en}}, \bibinfo
  {author} {\bibfnamefont {A.}~\bibnamefont {Simonsson}}, \bibinfo {author}
  {\bibfnamefont {A.}~\bibnamefont {K\"allberg}}, \bibinfo {author}
  {\bibfnamefont {P.}~\bibnamefont {L\"ofgren}}, \bibinfo {author}
  {\bibfnamefont {S.}~\bibnamefont {Mannervik}}, \bibinfo {author}
  {\bibfnamefont {H.}~\bibnamefont {Zettergren}}, \ and\ \bibinfo {author}
  {\bibfnamefont {H.}~\bibnamefont {Cederquist}},\ }\href {\doibase
  10.1103/PhysRevLett.119.073001} {\bibfield  {journal} {\bibinfo  {journal}
  {Phys. Rev. Lett.}\ }\textbf {\bibinfo {volume} {119}},\ \bibinfo {pages}
  {073001} (\bibinfo {year} {2017})}\BibitemShut {NoStop}%
\end{thebibliography}%

\end{document}